\documentclass[12pt]{article}     

\usepackage{graphicx}
\usepackage{amssymb}
\usepackage{amsmath}

\begin{document}

\title{Large gaps imputation in remote sensed imagery of the environment
}
\author{ Valeria Rulloni, Oscar Bustos and   Ana Georgina Flesia\thanks{AGF is corresponding
author }\\CIEM-Conicet and FaMAF-UNC\\C\'ordoba, Argentina\\ $\{vrulloni,bustos,flesia\}@famaf.unc.edu.ar$
}

\maketitle

\begin{abstract}

Imputation of missing data in large regions of satellite imagery is necessary when the acquired image has been damaged by shadows due to clouds, or information gaps produced by sensor failure.

 The general approach for imputation of missing data, that could not be considered missed at random, suggests the use of other available data. Previous work, like local linear histogram matching, take advantage of a co-registered older image obtained by the same sensor, yielding good results in filling homogeneous regions, but poor results if the scenes being combined have radical differences in target radiance due, for example, to the presence of sun glint or snow.

This study proposes three different alternatives for filling the data gaps. The first two involves merging radiometric information from a lower resolution image acquired at the same time, in the Fourier domain (Method A), and using linear regression (Method B). The third method consider segmentation as the main target of processing, and propose a method to fill the gaps in the map of classes, avoiding direct imputation (Method C). 

All the methods were compared by means of a large simulation study, evaluating performance with a multivariate response vector with four measures: Q, RMSE, Kappa and Overall Accuracy coefficients. 
Difference in performance were tested with a MANOVA mixed model design with two main effects, imputation method and type of lower resolution extra data, and a blocking third factor with a nested sub-factor, introduced by the real Landsat image and the sub-images  that were used. Method B proved to be the best for all criteria.

\end{abstract}

\section{Introduction}

Management of the environment and inventory of natural resources often requires appropriate remote sensing data acquired at specific times on earth locations. But quite often, good resolution optical images have large damaged areas with lost information due to clouds or shadows produced by clouds.

The general statistical approaches for imputation needs to consider the data loss in one of three categories: missing at random data,  completely missing at random data,  (meaning that the missing data is independent
of its value), and Non-ignorable missingness, Allison(2000). The last case is the most problematic form, existing when missing values are not randomly distributed across observations, but the probability of missingness cannot be predicted from the variables in the model. One approach to non-ignorable missingness is to impute values based on data otherwise external to the research design, as older images, Little et al (2002). 

Merging information acquired at the same time from two or more sensors  (with possible different resolution) is the core of  data fusion techniques. Their historical goals are to increase image resolution, sharpening or enhancement of the output,  Tsuda et al (2003); Ling (2007) and Pohl (1998), and their major problem to overcome is the co-registration of the different sources to merge, Blum et al  (2005).

 In the last decade several adaptations of data fusion techniques for  information recovery have been devised to mitigate the effect of clouds on optical images, a classical problem, since 50\% of the sky is usually covered by light clouds. 
  In Le Hegarat-Mascle et al. (1998) contextual modeling of information was introduced in order to merge data from SAR (Synthetic Aperture Radar) images into  optical images, to correct dark pixels due to clouds. Arellano (2003) used Wavelet transforms first for clouds detection, and then to correct the information of the located clouds pixels by merging older image information with a wavelet multiresolution fusion technique.  Rossi et al (2003), introduced a spatial technique, krigging interpolation,  for correction of shadows due to clouds. Shadows and light clouds do not destroy all information, only distort it, but dark clouds or rain clouds produce non-ignorable missingness, a harder problem for interpolation techniques.

\begin{figure}[h]
\centering
\includegraphics[width=5.1cm]{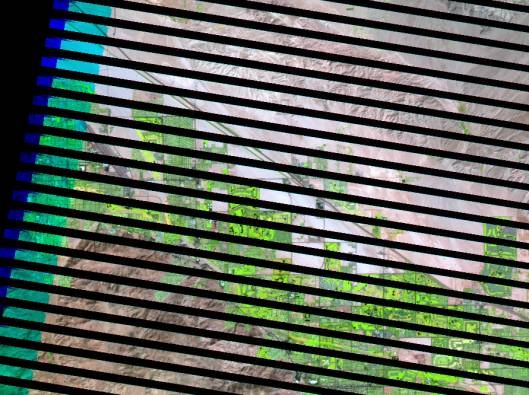}\;\;
\includegraphics[width=5.2cm]{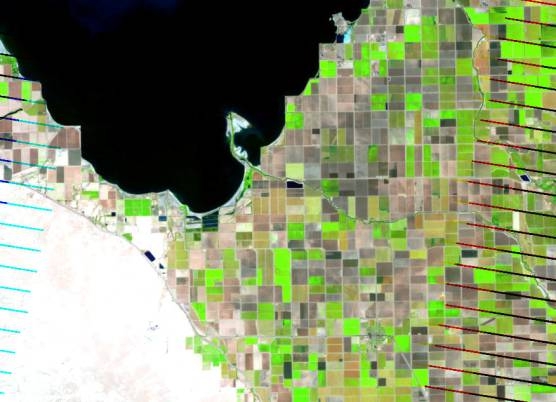}
\label{Zcent} \caption{Left panel, a piece of a Landsat 7 image with lost information. Right panel, a center piece of the same image almost without loss.}
      \end{figure}

Another interesting example of incomplete data is the damaged imagery provided by the Landsat 7 ETM+ satellite after its failure on May 2003. A failure of the Scan Line Corrector, a mirror that compensates for the forward motion of the satellite during data acquisition, resulted on overlaps of some parts of the images acquired thereafter, leaving large gaps, as large as 14 pixels, in others. About 22\% of the total image is missing data in each scene. In figure \ref{Zcent} we can see two parts of the same Landsat 7 image, one with missing information and another almost without loss.

The Landsat 7 Science Team (2003), developed and tested composite products made with the damaged Landsat 7 image and a database of  older Landsat images, using local linear histogram matching, a method that  merges pieces of older images that best matches the surroundings of the missing information region,  Scaramuzza et al 2004, USGS/NASA 2004, (United States Geological Survey/ National Aeronautics and Space Administration). Commonwealth of Australia (2006) reported that the composite products appear similar in quality to non damaged Landsat 7 images, but masked  environment changes, an usual problem in data fusion when one of the sources is temporally inaccurate.

Zhang and Travis (2006) developed a spatial interpolation method based on krigging, the krigging geostatistical technique, for filling the data gaps in Landsat 7 ETM+ imagery, without need of extra information. They compare their method with the standard local linear histogram matching technique chosen by USGS/NASA. They show in their case study that krigging is better than histogram matching in targets with radical differences in radiance, since it produces an imputation value closer to the values of the actual neighbors. A drawback of spatial techniques is that they rely on a neighborhood that could be completely empty of real information. The algorithms start imputation in the gap contour, where there are many good pixels, and use imputed and good values to create the next pixel value. In center pixels of large gaps, algorithms based only on damaged images will use previously imputed values to generate the next imputation, degrading visual quality and increasing interpolation error.

In this paper, we propose three methods based on data fusion techniques for imputation of missing values in images with non-ignorable missingness. We suppose there is available temporally accurate extra information for the gap scenes, produced by a lower resolution sensor. This is not a very restrictive hypothesis, since there are many satellite constellations that can provide temporally accurate data with different cameras at different resolutions. 

This study proposes three different alternatives for filling the data gaps. The first method involve merging information from a co-registered older image and a  lower resolution image acquired at the same time, in the Fourier domain (Method A). The second used a linear regression model with the lower resolution image as extra data (Method B). The third method consider segmentation as the main target of processing, and propose a method to fill the gaps in the map of classes, avoiding direct imputation for the segmentation task (Method C). Radiometric imputation is later made assigning a random value from the convex hull made by the neighbor pixels within its class. 

All the methods were compared by means of a large simulation study, evaluating performance with a multivariate response vector with four measures: Q, RMSE, Kappa and Overall Accuracy coefficients. Two of these performance measures (Kappa and Overall Accuracy) are designed  for the assessment of segmentation accuracy. The other two, Q and RMSE, measure radiometric interpolation accuracy.  
Difference in performance were tested with a MANOVA mixed model design with two main effects, imputation method and type of lower resolution extra data, and a blocking third factor with a nested sub-factor, introduced by the real Landsat image and the sub-images  that were used. Method B proved to be the best for all criteria.

\section{Methods\label{fun imag}}

We are considering images acquired from the same geographic target, with fixed number of bands $K$, and 256 levels of gray per band. Damaged and older images have the same support, lower resolution images have different supports, and we suppose that there are $n_z$ pixels in the damaged image for each pixel in the lower resolution image.

Let  $X_D$ be the damaged image,  $Z$ a lower spatial resolution image , and  $X_{old}$, an older image acquired with the same sensor that $X_D$. Let $S_D$ be the gap to be filled, i.e. the  set of pixels of $X_D$ with missing values. The goal is to input values on $S_D$ using available good data from $X_D$ in a small neighborhood of each pixel, the values of $Z$ and eventually, the values of $X_{old}$.

The first step in processing is the re-sampling of the lower resolution image in order to match the support of the three images. This is done replacing each pixel $(c,r)$ of $Z$ by a matrix of size $n_z\times n_z$ with constant value $Z(c,r)$.

Imputation is done in each band separately with the same algorithm, thus the methods descriptions consider the images as one band only.

\subsection{Method A}

The image $X_{old}$ does not need to be co-registered with reference to the damaged image $X_D$, since they have been acquired by the same sensor, but calibration is indeed necessary to increase merging accuracy.

We suppose a Gaussian calibration have been made on $X_{old}$ per column, taking as pivots the values in the matched column of $X_D$. Let $X_R^A$ be the composite image, output from the imputation method A.  We define $X_R^A$ values in the gap as a mixture of high frequencies of the older image and low frequencies taken from the actual but lower resolution image.

\[
X_R^A(c,r)=\left\{ \begin{array}{lr}\left\vert L_{C}(Z)(c,r)+H_{C}(X_{old})(c,r)\right\vert & (c,r)\in S_D\\
 X_D(c,r) & \mbox{otherwise}
\end{array}\right.
\]
with $H_C$ and $L_C$ the ideal High pass and Low Pass Fourier Filters.

The filters are regulated by the $C$ coefficient, the bigger $C$ is the larger the influence of $Z$ in the imputation. High pass filters are related to detail, edges and random noise, and low pass filters to structural information. Therefore, Method A takes structural information from the low resolution image and detail from the older one.

\subsection{Method B}

Imputation will be made with the only help of $Z$, a temporally accurate image with lower resolution. We have expanded the lower resolution image to match the support of $X_D$, replacing each cross grained pixel by a block of  $n_z\times n_z$ pixels with the same radiometric value. Each one of the constant blocks of the expanded $Z$ image has a matched block in the damaged image $X_D$, that could be valid (having all the information), or non valid, (with some loss).

We will follow a time series approach now. Let think we have data collected month to month along several years, and we have a missing January data. It is reasonably to impute that value using a regression model that only involves other January data, and extra data collected on Summer that year.

In our case, we have missing information on a location $(c,r)$ inside a block $B$, we may impute that value using a regression model that only involves data in the same position inside the other blocks of the image (January data), using as regressor variable the values of $Z$ (Summer data).
\begin{equation*}
X_D(c,r)=\alpha_{(c,r)}*Z(c,r, B)+\beta_{(c,r)}+\varepsilon
(c,r),
\end{equation*}
with $\varepsilon(c,r)\sim N(0,\sigma(c,r)^{2})$.

The coefficients $\alpha_{(c,r)}$ and
$\beta_{(c,r)}$ are estimated by ordinary least squares using only valid blocks. Then
\begin{equation*}
\widehat{\alpha }(c,r)=\frac{\displaystyle\sum_{B
{\mbox { a valid block}}}(Z(c,r,B)-\overline{Z}%
)(X_D(c,r,B)-\overline{X_D(c,r)})}{\displaystyle\sum_{B
{\mbox { a valid block}} }(Z(c,r,B)-\overline{Z})^{2}},
\end{equation*}

\

\begin{equation*}
\widehat{\beta }_{(c,r)}=\overline{X_D(c,r)}-\widehat{\alpha }_{(c,r)}%
\overline{Z},
\end{equation*}

\

where
\begin{equation*}
\overline{Z}=\sum_{{B
{\mbox { a valid block}}}}\frac{Z(c,r,B)}{%
\left| {{\mbox {number of valid blocks}}}\right| }
\end{equation*}
and
\begin{equation*}
\overline{X_D(c,r)}=\sum_{{B
{\mbox { a valid block}}}}\frac{X_D(c,r,B)%
}{\left| \mbox {number of valid blocks}\right| }.
\end{equation*}

\

We define the imputed image $X_R^B$ as  $X_D$ in the pixels with no missing information and the value predicted by the regression in each damaged pixel .

\begin{equation*}
X_R^{B}((d,g,B))=\left\{
\begin{array}{lr}
\widehat{\alpha }(d,g)\ast Z(d,g,B)+\widehat{\beta }(d,g) &\ \ (d,g)\in S_{D} \\
X_D((d,g,B))& \mbox{otherwise}
\end{array}
\right. .
\end{equation*}

\subsection{Method C}

Remote sense images are used in a wide range of applied sciences. For many of them, as Agricultural and Experimental Biology, or Environmental Science, all useful information is contained on a class map, where the classes are characterized by special features under study. Kind of crop, forested or deforested areas, regions with high, medium and low soil humidity, are examples of such features. 

 In this section we will change our point of view by thinking on class maps instead on radiometric images. Let suppose we have our damaged image $X_D$, and a temporally accurate image $Z$ with lower resolution, and possible different number of radiometric bands.  This is an advantage over radiometric imputation, which need the same spectral properties in both images.

We also suppose that a map of classes $C_D$ has been drawn from the damaged image $X_D$, using a non supervised method like $K$ means, with $K$ different classes. The class of missing information pixels is another class, called class 0. The goal is to assign the pixels without radiometric information to one of the $K$ classes, and generate a radiometric value for them randomly from the selected class.

There are two main steps in this method

\begin{enumerate}

\item An enhancement of the initial classification $C_D$, that could be made or not

\item Imputation of pixels in the zero class.

\end{enumerate}

\paragraph{Initial enhancement}

\

Automatic classification is a difficult task, and the imputation method under study heavily depends of the accuracy of the initial map of classes. So it is important to pay special attention to the coherence of the classes. We suggest a set of steps to improve class homogeneity as follows

\begin{enumerate}
\item Given $X_D$ construct a map of classes $C_D$ using $K$ means, and assign the zero label to the missing information regions.
\item Given the class image $C_D$, detect the pixels with  non homogeneous neighborhood, i.e,  pixels that have no neighbor pixels on the same class. Call this set $N_G$.
    \item For each pixel in $N_G$, verify its label using a forward mode filter, and a backwards mode filter. The filters with give a label that is consistent with the mode of the labels in a (forward or backward) neighborhood. 
        \begin{enumerate}
        \item If both filters give the same label, update the label of the pixel to match this one.
        \item If the labels are not the same, maintain the original label, and put the pixel in a new set call $N_M$.
        \end{enumerate}
        \item For each pixel in $N_M$ that do not has label zero, we use again radiometric information for updating the label. 
        \begin{enumerate}
        \item Take the one step neighbors of the pixels, and compute the arithmetic mean for each class present in the neighborhood. 
        \item Update the label of the pixel by the label of the class whose arithmetic mean is closest (in Euclidean distance) to the radiometric value of the pixel.
            \end{enumerate}
\end{enumerate}

It is important to note that after this process, some of the pixels of the zero class may have been classified into a real class, only using contextual information. The other classes should have more defined borderlines.

\paragraph{Class zero imputation}

\

We suppose now that we have a stable map of $K$ classes $C_D$, with missing information in class zero, and auxiliary information provided by a temporally accurate lower resolution radiometric image $Z$.
If $(c,r)$ is a pixel in class zero, it does not have radiometric information. We will impute a label class on it with the following algorithm
 
 \begin{enumerate}
 \item Let $N_t$ be the smallest neighborhood of $(c,r)$  that have a pixel with non zero label.
  
  \item Let $\overline Z_k$ be the arithmetic mean of the $Z$ pixels that have positions in class $k$ in $N_t$.
  \item The label of $(c,r)$ is the label of the class $k$ that makes the smallest Euclidean distance from $Z(c,r)$ to each $\overline Z_k$. 
\[
k=\mbox{arg} \min_{k} \left\Vert Z(c,r)-\overline{Z}_{k}(c,r)\right\Vert _{2}.
\]
\end{enumerate}

After this process, all pixels have a class label, but pixels in the gaps have not been imputed with radiometric information yet. A pixel's  radiometric value is assigned  randomly from the values of the convex hull generated by the pixels of a small neighborhood within its class. 

\section{Precision assessment by simulation}

Impartial imputation assessment is only possible when ground truth is available. Complete simulation of the three types of images involved in the methods (old, damaged and lower resolution) will introduce errors beyond the ones produced by the imputation methods, degrading the quality of the assessment.
For this reason,  good quality Landsat 7 ETM+ imagery that have older matched imagery available were selected, and  strips similar to the gaps in the SLC-off ETM+ imagery were cut manually, guaranteeing ground truth to compare with and co-registration between the images. 
The four Landsat 7 ETM+ images selected were quit large, having many different textures in them, like crop fields, mountains and cities, that challenge the imputation methods differently. 

Landsat 7 ETM+ had a lower resolution sensor in its constellation, the MMRS camera from the SAC C satellite, whose imagery could be used as extra data. But to control also co-registration problems, lower resolution imagery were simulated with three resolution reduction methods (RRM),  by block averaging the ground truth, the  CONGRID method from ENVI software and shifted block averaging. Block averaging is a crude way of reducing resolution, the CONGRID method reduces the blocking effect smoothing the output, giving better visual appearance, but block averaging allows to shift the blocks in a controlled fashion, simulating lack of co-registration.  

\begin{figure}[h]
\centering
\includegraphics*[width=5.6cm]{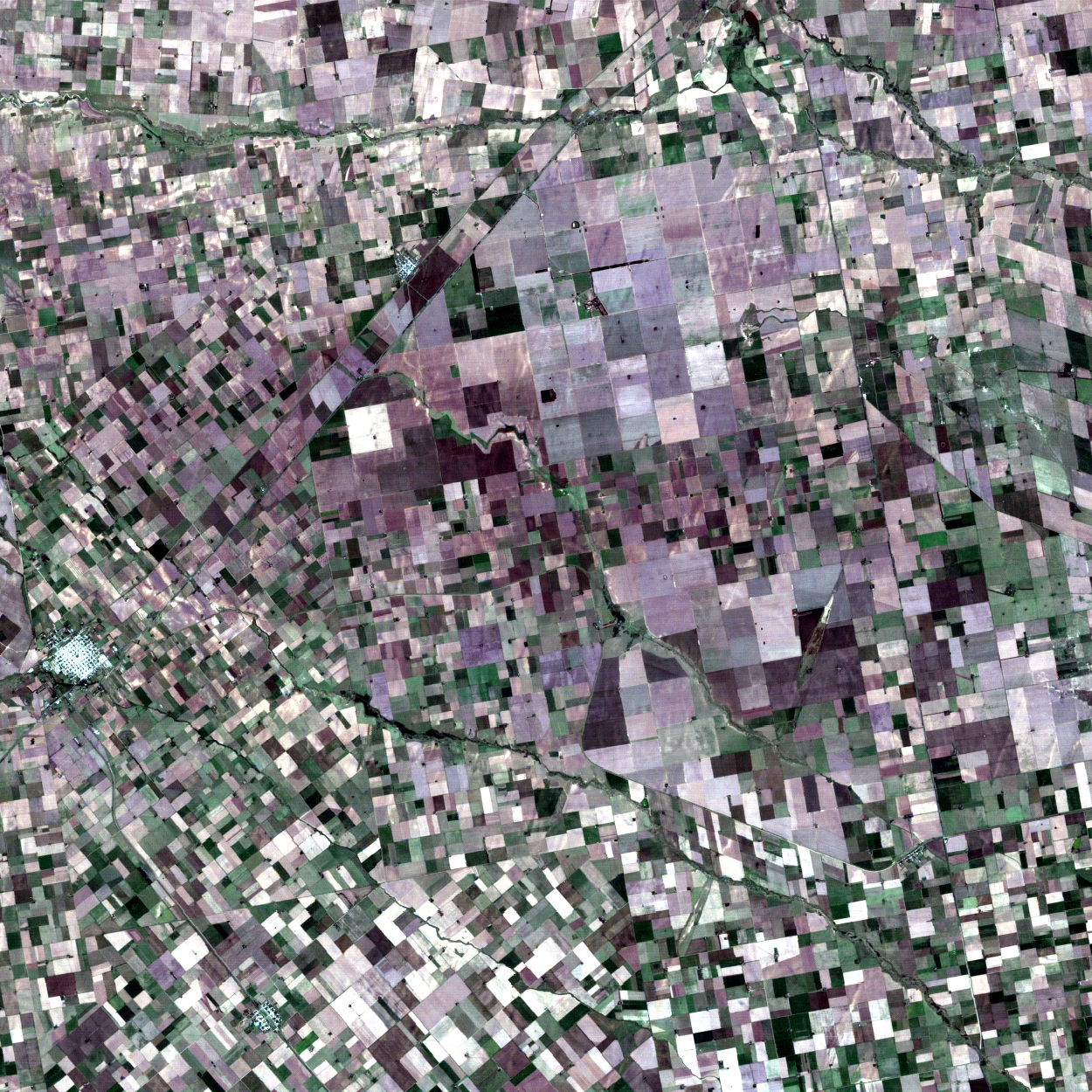}\;
\includegraphics*[width=5.6cm]{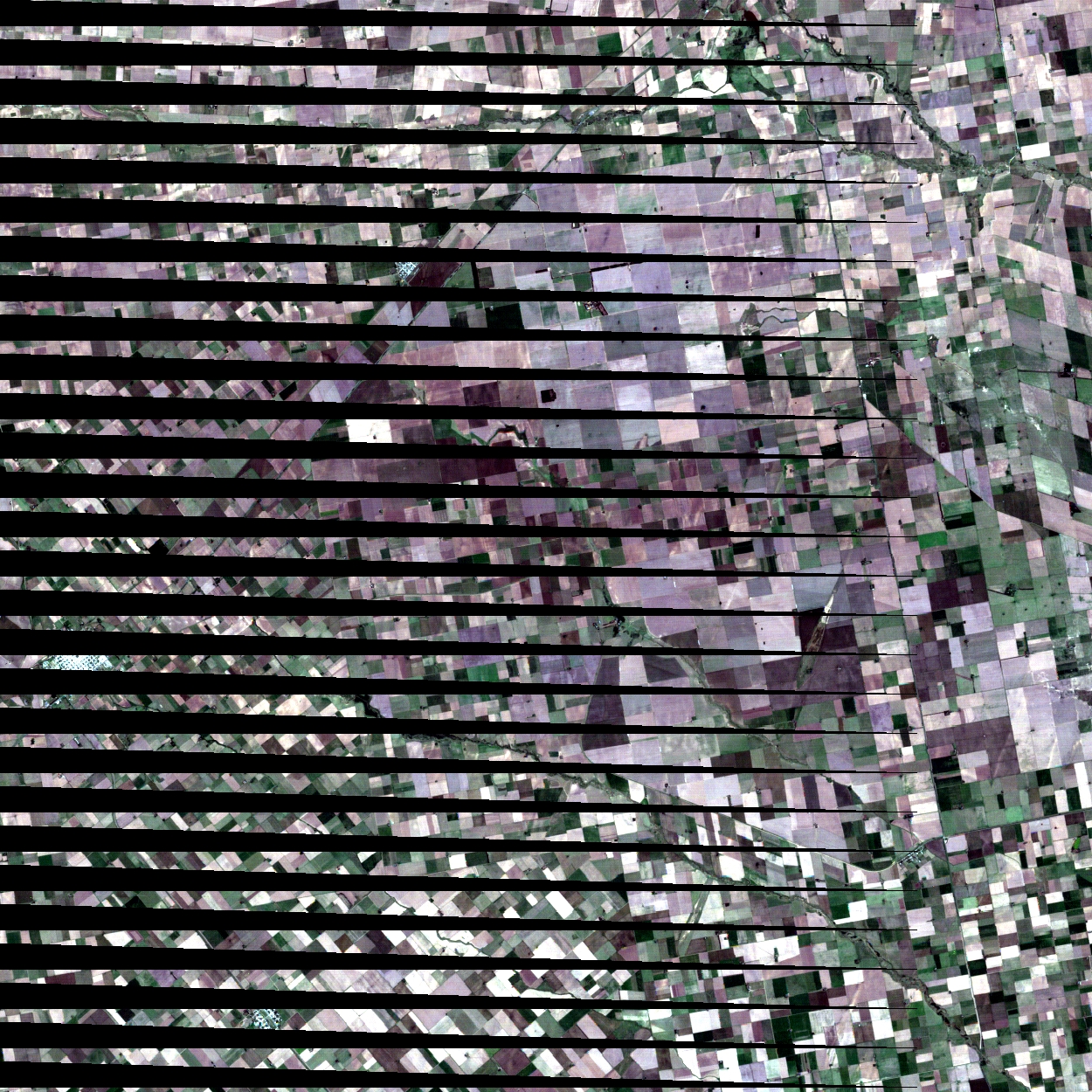}\vspace{0.1cm}
\includegraphics*[width=5.6cm]{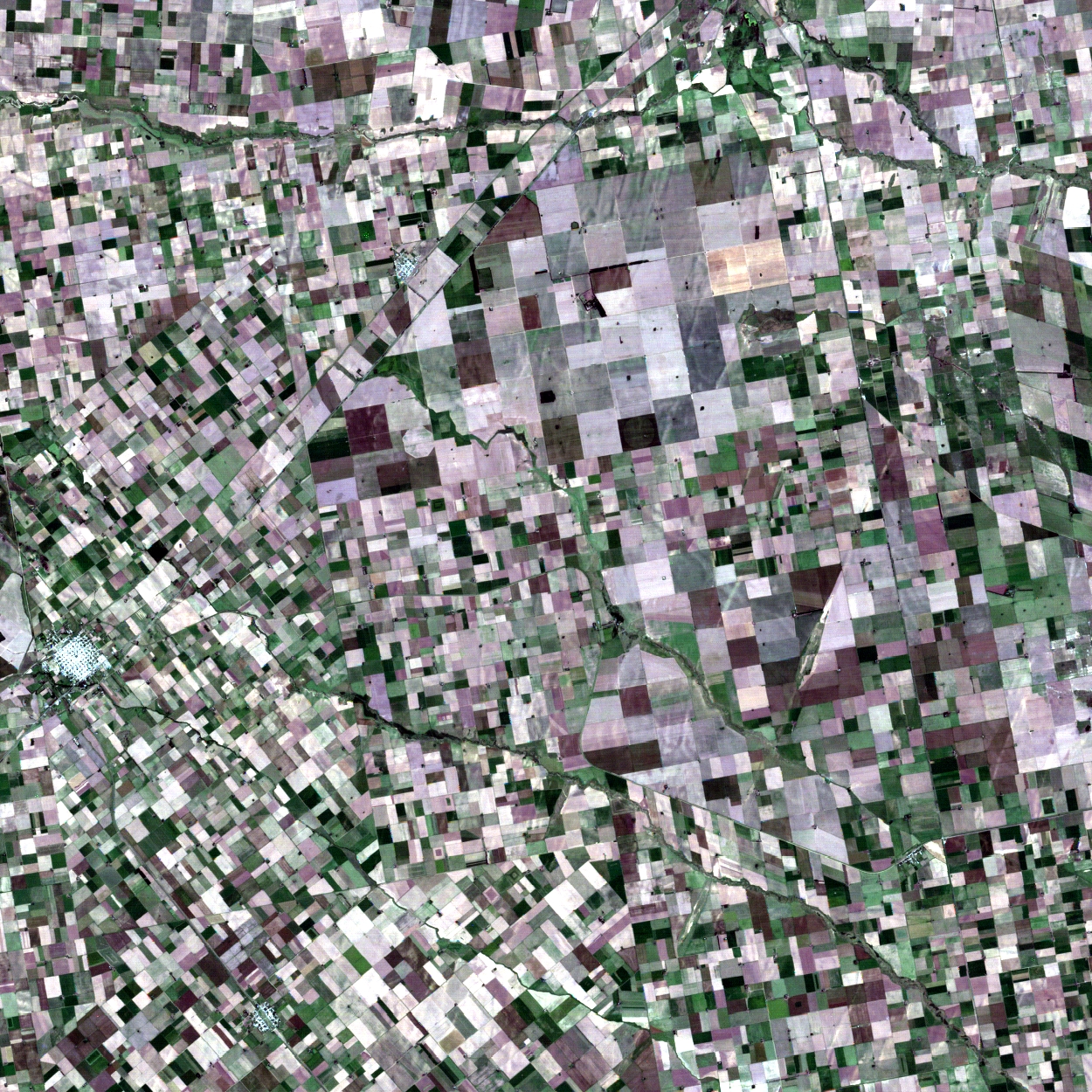}\;
\includegraphics*[width=5.6cm]{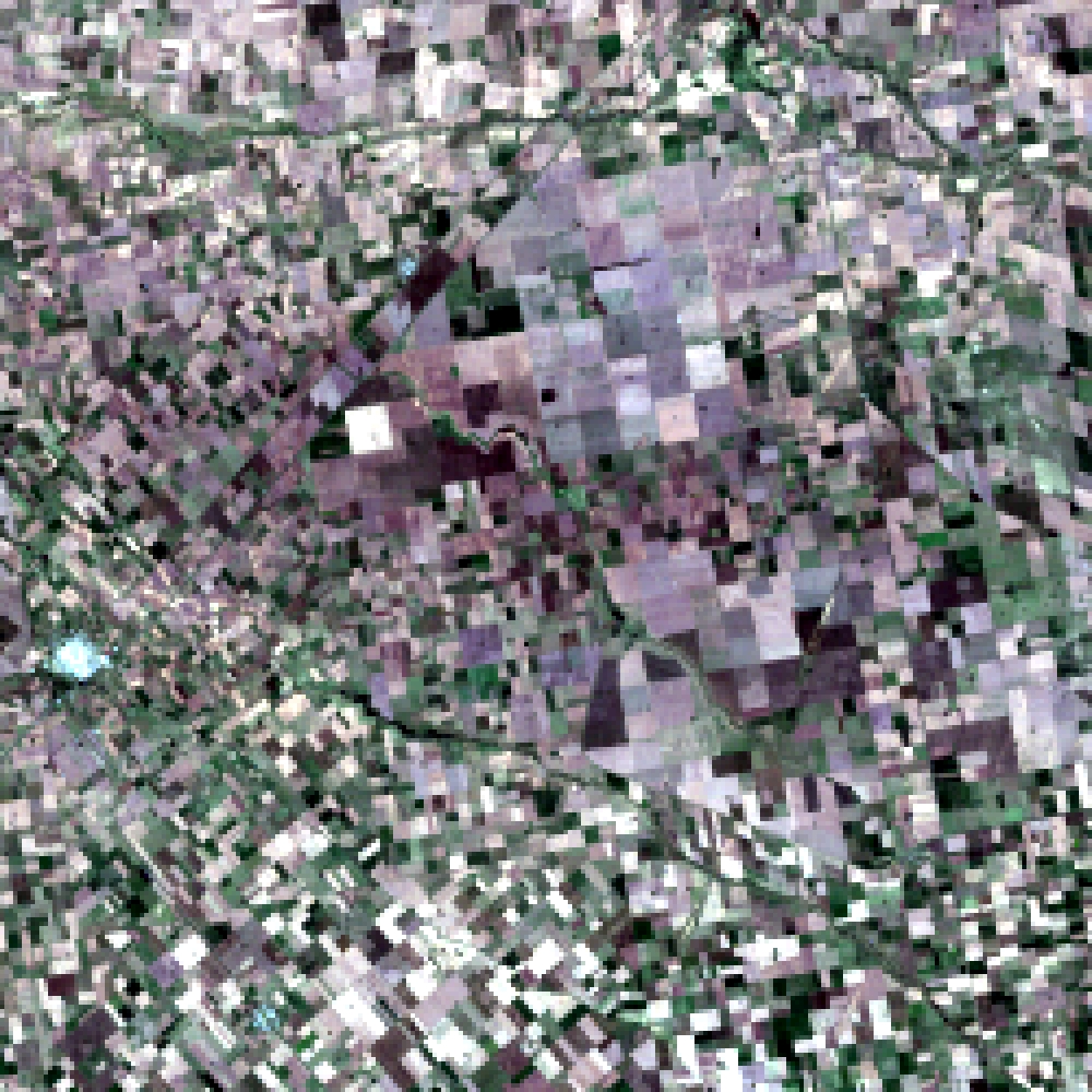}
 \caption{Examples of images from database. From left to right, top to bottom, sample of Landsat 7 image, damaged image, older image and lower resolution image}
\label{examples}
\end{figure}

 In Figure \ref{examples} we see an example of four matched images, good, damaged, older and lower resolution by block averaging. We can see a bright spot in the center of the image, which is still present in the damaged one and the lower resolution one, but it is not present in the older one.  Database construction details are giving in the  subsection \ref{database}.

The performance measures (RMSE, Q, Kappa and Overall Accuracy) were modeled as a response  vector in a MANOVA  mixed model with main effects (imputation method, resolution reduction method), and random nested effects (image and sub-image). Subsection \ref{measures} introduces the performance measures and subsection \ref{anova} includes a detailed description of the MANOVA  design. Following the Manova rejection, simultaneous multivariate comparisons were made by (Bonferroni corrected) two means Hotelling tests, and individual Anova results with simultaneous comparisons via Fisher LSD tests  were studied to determine the best method, and the influence in performance of  co-registration problems.

\subsection{Database}
\label{database}

We selected four Landsat 7 ETM+ images with six bands each,  acquired before the failure of the SLC mirror. These images have also companion good quality Landsat images acquired approximately an year before. In the Table~\ref{tab}, we see some information about the images selected, courtesy of CONAE-Argentina.
\begin{table}[h]
\caption{Landsat 7 ETM+ Images selected for simulation study, courtesy of CONAE-Argentina.}\label{tab}
\begin{tabular}{|c|c|c|c|}
\hline
Track & Frame &  L7 image date & old L7 image date \\ \hline
1 & 73 & 02-MAY-2003\_14:24:38.069 & 29-APR-2002\_14:24:42.108 \\ \hline
1 & 74 & 02-MAY-2003\_14:25:02.021 & 29-APR-2002\_14:25:06.058 \\ 
229 & 83 & 04-NOV-2002\_13:56:49.481 & 17-NOV-2001\_13:57:29.017 \\ \hline
229 & 87 & 04-NOV-2002\_13:58:25.329 & 01-NOV-2001\_13:58:52.526 \\ \hline
\end{tabular}
\end{table}
A data set of small images for experimenting was constructed in a hierarchical way, in order to account for variability between images acquired at different times and locations. Four pieces of each different Landsat 7 ETM image were selected from regions with few clouds and then subdivided again in four pieces, obtaining 16 sub-images for each one of the original set of Landsat Images.   
The final 64 sub-images have 1250 $\times$ 1250 pixels (and 6 bands), and were stored as ground truth.
Their companion damaged images were computed erasing  26\% of information of each image, mimicking the strips with missing information in the SLC-off Landsat imagery.

It is well known that resolution in optical images can be degraded by block averaging. Using in the study a truly lower resolution image like the one provided by the SAC C satellite camera, would have introduced extra calibration and co-registration errors, without control over them. It was decided then to degrade the ground truth Landsat image by averaging, and simulate lack of co-registration shifting the degraded image a few lines and columns. This procedure will make possible the assessment of the impact of lack of co-registration in the imputation methods. 

For each one of the 64 images, three types of lower resolution images  with the same support were generated
\begin{enumerate}
\item $Z_0$, averaging radiometric information of  5$\times$5 pixels square regions. Image was divided into 5$\times$5 consecutive blocks, redefining the radiometric value of each pixel within the block as the average value of radiometric information of the original pixels 
\begin{equation*}
Z_0(ic,ir,B)=\frac{1}{\text{5*5}}\sum_{c=0}^{\text{5-1}%
}\sum_{r=0}^{\text{5-1}}\text{L7}(c,r,B))
\end{equation*}
$\forall (ic,ir)\in B$, 5$\times$5 block.
\item $Z_1$ interpolation given by the IDL  CONGRID command that reduces the impact of blocking by smoothing the output.
\item $Z_2$ moving $Z_0$ 3 rows and 2 columns, imitating a possible small co-registration problem.
\end{enumerate}
The companion older Landsat images were also cropped to match the sub-figures design. In Figure \ref{examples} we have shown a sub-image with many agricultural fields and city blocks, and the matched sub-image, taken approximatively one year before. Main structure is the same, roads, ponds, main agricultural fields, but fine scale structure is different, and there is a possible change in crops, since homogeneous patches look different.

\subsection{Performance measures}
\label{measures}
Radiometric evaluation and comparison were made globally and locally by means of  RMSE (Root mean square error) and  Q coefficients . The differences between the map of classes were assessed with measures related to the confusion matrix,  Overall Accuracy and Kappa coefficients. This section gives the definition of all four measures. Further information about radiometric comparisons can be found for instance,  in Congalton et al (1991) and Richards et al (1999).

\subsubsection{Radiometric measures}

\paragraph{RMSE}
The Root Mean Square Error between two images $Y$ and $W$ in region $S_0$ is 
\begin{equation*}
\sqrt{\frac{\sum_{s\in S_{0}} \left\| Y_{s}-W_{s}\right\|
^{2}}{\left\vert S_{0}\right\vert }}
\end{equation*}
The values of the RMSE are scale and order dependent, thus there is not universal reference values to compare with. Simulating the damage in the images, the scale and order are preserved, so the RMSE is not contaminated with distortion noise.

\paragraph{Q-coefficient}
Wang and  Bovik (2002) propose $Q$ as an objective index of image quality  that measure distortions with a combination of three factors: correlation loss, illumination distortion and contrast deformation .
Let $I$ be
\begin{equation*}
I=\frac{4\sigma _{YW}\overline{Y}\overline{W}}{\left( \sigma
_{Y}^{2}+\sigma _{W}^{2}\right) \left[
\overline{Y}^{2}+\overline{W}^{2}\right] }
\end{equation*}
where\newline
\begin{equation*}
\overline{Y}=\frac{1}{N}\sum_{i=1}^{N}Y_{i}, \sigma _{Y}^{2}=\sigma
_{YY}, \sigma _{YW}=\frac{1}{N-1}\sum_{i=1}^{N}\left(
Y_{i}-\overline{Y}\right) \left( W_{i}-\overline{W}\right).
\end{equation*}

The image is divided in  $M$ small regions, and the index $I$ is calculated in each one. The overall quality index $Q$ is the average over all them.

\begin{equation*}
Q=\frac{1}{M}\sum_{j=1}^{M}Q_{j}
\end{equation*}

The $Q$ range is $\left[ -1,1\right] $, the highest value is obtained when  $x_{i}=y_{i}$ for all $i=1,...,N$.

\subsubsection{Confusion matrix: measures for class accuracy evaluation}

Let $M$ be the standard confusion matrix generated by two map of classes,  the original map of classes $C_X$ and the reconstructed $C_D$. The matrix 's $M(i,j)$ entry is   $n_{\text{ij}}$, the number of pixels  of the true class $j$ assigned to class $i$  
in the reconstructed image.

The totals by column and row will be denoted by ($n_{\text{+j}}$) and ($n_{\text{i+}}$) ,  $n$ will be  the total of pixels considered and $p_{\text{ij}}=n_\text{ij}/n, p_{i+}=n_\text{i+}/n, p_{+i}=n_\text{+j}/n$ the proportions of misclassified pixels.

\paragraph {Overall Accuracy}

\begin{equation*}
OA=\frac{\sum_{i=1}^{\text{k}}n_{ii}}{n}
\end{equation*}
This coefficient compute the proportion of coincidences between the reference and reconstructed images (well classified pixels). A classification scheme is considered good if its Overall Accuracy is above 85\%.

\paragraph{ Cohen's Kappa coefficient}

\begin{equation*}
\widehat{K}=\frac{p_{o}-p_{c}}{1-p_{c}}.
\end{equation*}
This coefficient is based in the real match
\[p_{o}=\sum_{i=1}^{k}p_{ii}\] and the possible match\[p_{c}=\sum_{i=1}^{k}p_{i+}p_{+i}\] It is an estimator of the degree of matching in a classification scheme, it helps determine if the scheme output is better than  random allocation. For further details see par example Agresti (1996).

\subsection{Imputation methods}

The database contains 64  sets with 6 images each. A 6-uple has one ground truth image $X$, one matched  damaged image $X_D$, one matched older image $X_{old}$ and three lower resolution images $Z_0, Z_1, Z_2$, outputs of the Resolution Reduction methods. The ground truth of each one of the 64 sets is a Landsat image of a different geographic location, with a wide range of textures. Cities, forests, mountains and agricultural fields can be found in each of these images.

Six classes of imputation methods were implemented in IDL: Method A with 3 radios for the High pass and Low pass filters, 20,50,80 \% ( A1,A2,and A3), method B, and Method C with and without classification enhancement (C and C1). 
All these methods were applied to the damaged image in each set, and the performance measures collected.

\subsection{ Mixed model Analysis of a Randomized Complete Block Design}

\label{anova}
We want to assess the main effects in the mean of the response vector of four performance measures induced by two factors:
\begin{itemize}
\item Imputation Method, with 6 levels : A1, A2, A3, B, C y C1

\item Resolution reduction method, with 3 levels: 0, 1 and 2;
\end{itemize}

Considering our set of Landsat images as a random subset of all possible images, we set image as random factor, and  sub-image of each image as a nested random factor. 

Our multivariate model design is 

\begin{eqnarray*}
 Y_{ijkl}&=&\mu +\alpha _{i}+\gamma _{j}+m_{k}+s(m)_{kl}+\varepsilon _{ijkl}\\
i &=&1,..,6 , \ j =1,..,3, \; k=1,..,4, \; l =1,..,16
\end{eqnarray*}
where
\begin{itemize}
\item  $Y_{ijkl}$ is the response vector RMSE, Q, Kappa and Overall Accuracy,
\item  $\alpha _{i}$ is the main effect of the imputation method,
\item  $\gamma _{j}$ is the main effect the resolution reduction method, 
\item $\mu $ is the general mean; 
\item $m_{k}$ and $s(m)_{kl}$ are the independent random effects introduced by images and sub-images,  $m_{k}\sim N(0,\Sigma_m)$,  $s(m)_{kl}\sim N(0,\Sigma_s)$ 
\item $\varepsilon _{ijkl}$ the independent experimental error: $\varepsilon _{ijkl}\sim N(0,\Sigma_e)$
\end{itemize}
All methods have been applied to all the sub-images, and no interactions have been considered since each image is a block itself and has no interesting interaction with the any of the methods.

Then
\begin{equation*}
Y_{ijkl}\sim N(\mu +\alpha _{i}+\gamma _{j},\Sigma _{s}+\Sigma
_{m}+\Sigma _{e})
\end{equation*}

The Manova model can be used to test the hypothesis that the vector of performance measures for each level of the main factors, imputation method and lower resolution method,  are equal against the alternative hypothesis that at least one is different. If the null hypothesis is rejected, numerous procedures can be used to determine which of the main effects is significantly different form the others. The comparison-wise Type I error will be fixed at $\alpha=0.05$, in order to give $F$-values, but we will report the $p$-values also in all cases . Following a Manova rejection, it have been seen that the overall experiment-wise error rate stays within the limits of the $\alpha$ selected for individual Anova tests, helping answering the following research questions

\begin{enumerate}

\item
Is there at least one method of imputation with a different performance from the others, within each performance measure analysis?

\item
Is there at least one method of resolution reduction with a performance different from the others, within each performance measure analysis?

\item Does comparing performance (on any factor level), relative to the different measures of performance, produce the same results?

\end{enumerate}

\section{Results}

\subsection{Empirical findings}

\subsubsection{Visual Inspection}
 \begin{figure}[ht]
\centering
\includegraphics*[width=5cm]{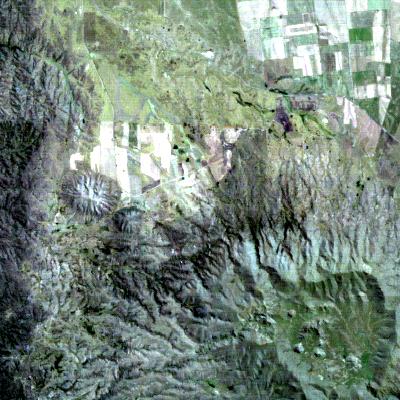}\;\includegraphics*[width=5cm]{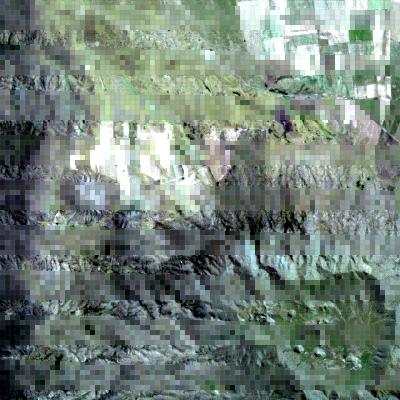}\vspace{0.1cm}
\includegraphics*[width=5cm]{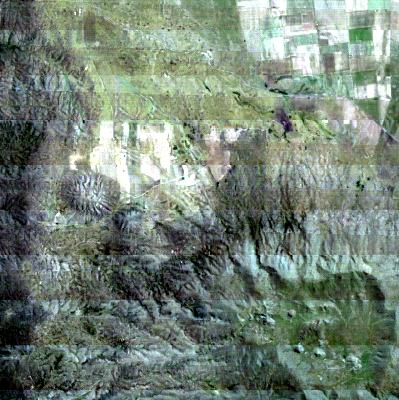}\;\includegraphics*[width=5cm]{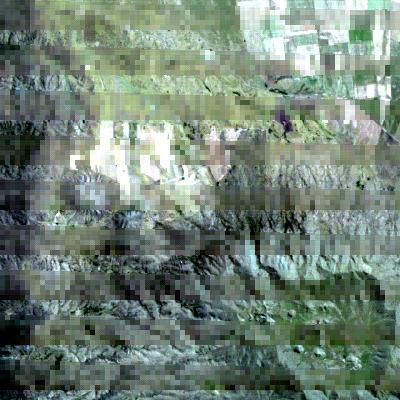}\vspace{0.1cm}
\includegraphics*[width=5cm]{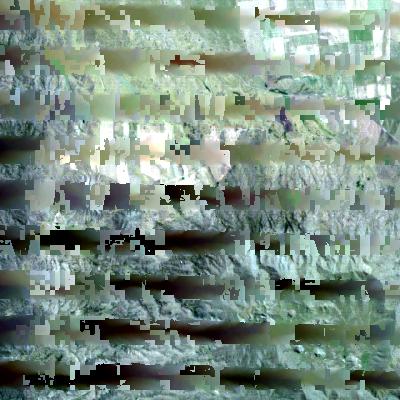}\;\includegraphics*[width=5cm]{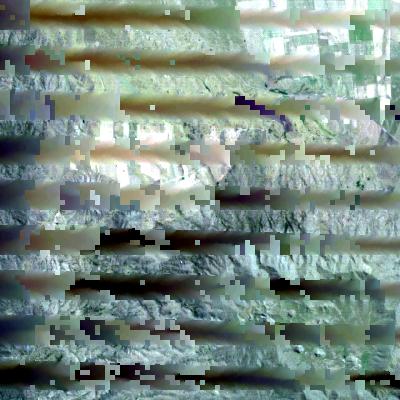}
 \caption{From left to right. First row: Original Sub-Image, reconstructions from Method B. Second Row: Reconstructions from Method A1, and Method A3.Third Row: Reconstructions from Method C and Method C1 }
\label{method}
      \end{figure}

In Figure \ref{method} it can be seen the original image and its reconstructions using methods B, A1 and A3, C and C1. We should notice that radiometric imputation from the map of classes are the worse of the methods, texture and fine details are not reconstructed, giving a blurred appearance to the imputed stripes. But we stress that it is not that bad if the goal is classification.

We should also notice that globally, Method B produces an output with better visual quality than the others. Stripes are missing, but fine lines, small scale structure is less defined, like blurred. The reconstructions made with Methods A1 and A3 show changes between the zones that are original and the zones imputed, but the imputed stripes have better defined small scale structure. This is particularly noticeable in reconstruction with Method A1, which introduces more information from the older image than Method A3. 

\subsubsection{Means Profiles}\label{profiles}
Visual inspection can only be made with few images, we have 64 sub-images with double size from the ones just shown. To inspect performance globally,  we will study now the mean profiles of all performance measures. We compute the sample means of each performance measure, for each method and each image. This is, we average the values of a particular performance measure taken over all 16 sub-images from a single image. A Mean Profile is a polygonal that links the four means (one per image) from a particular performance measure, .
 
 Methods mean profiles should be parallel, if there is no interaction between the performance measure and the images. Differences are then produced by the methods and not by the images. Flat Profiles indicates that each method produce similar values on any image, but that is not true. Images depend not only of the geographic target observed, but the atmospheric conditions of the moment of acquisition. Variability is huge. If we consider our images as chosen at random, they will be only a random effect that inflates the variance, but does not introduce a main effect that could mask the effect of the Imputation Methods.

Also, it is important to note that in order to compare all methods using Kappa and Overall Accuracy, a map of classes must be made for each reconstructed image, as well as the original image. Methods A and B generates radiometric imputations, Method C and C1 generate maps of classes first, and then radiometric imputation, choosing at random a value within the class assigned. The automatic method $K$-means was used to obtain a map of classes for the original images and the ones reconstructed with methods A and B.

The coefficients Kappa, $Q$ and Overall Accuracy are designed to give high scores to good reconstructions. On the contrary, RMSE is an error measure, then reconstruction is best when RMSE is at its lowest.

\begin{figure}[h]
\centering
\includegraphics*[width=7.5cm]{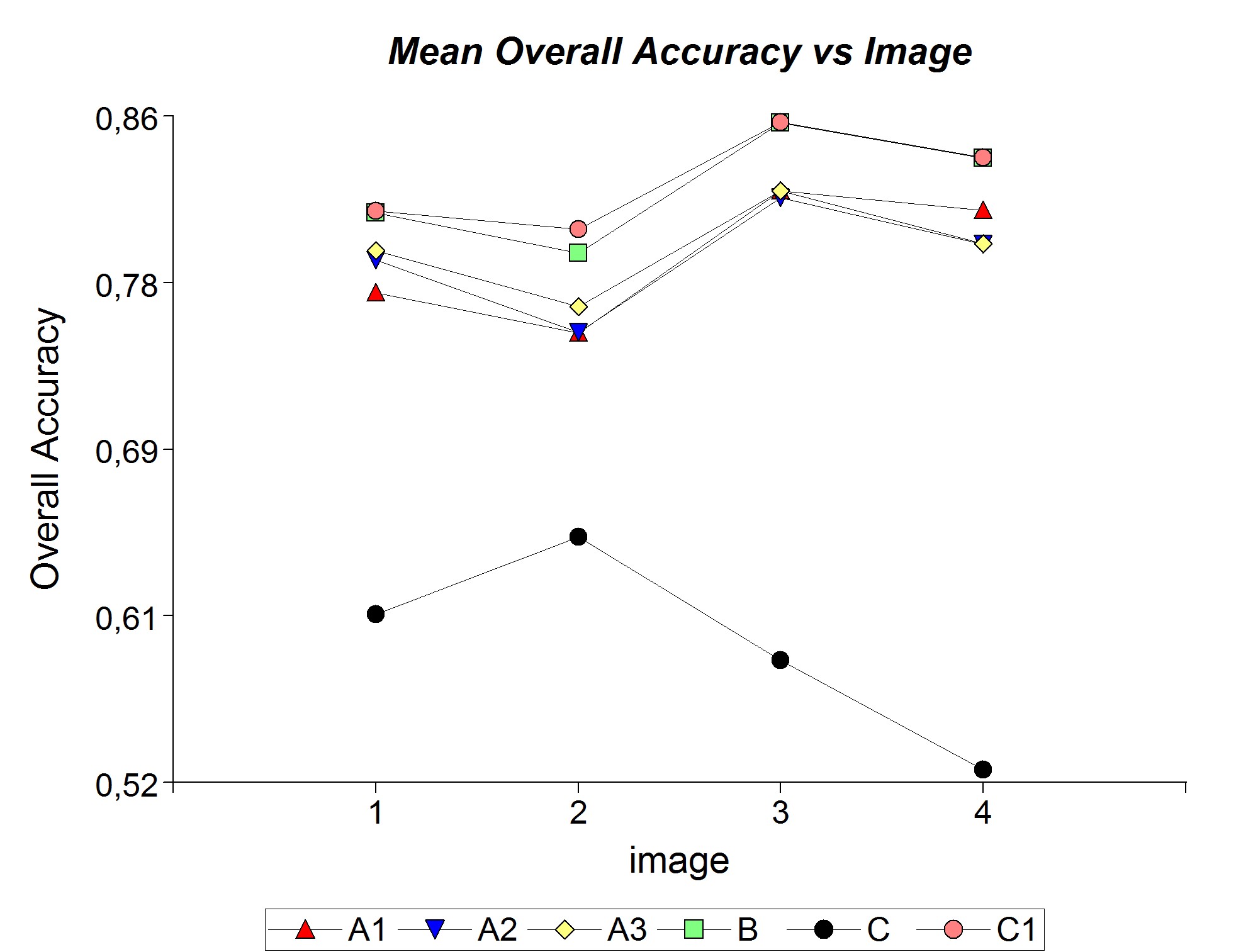} \includegraphics*[width=7.5cm]{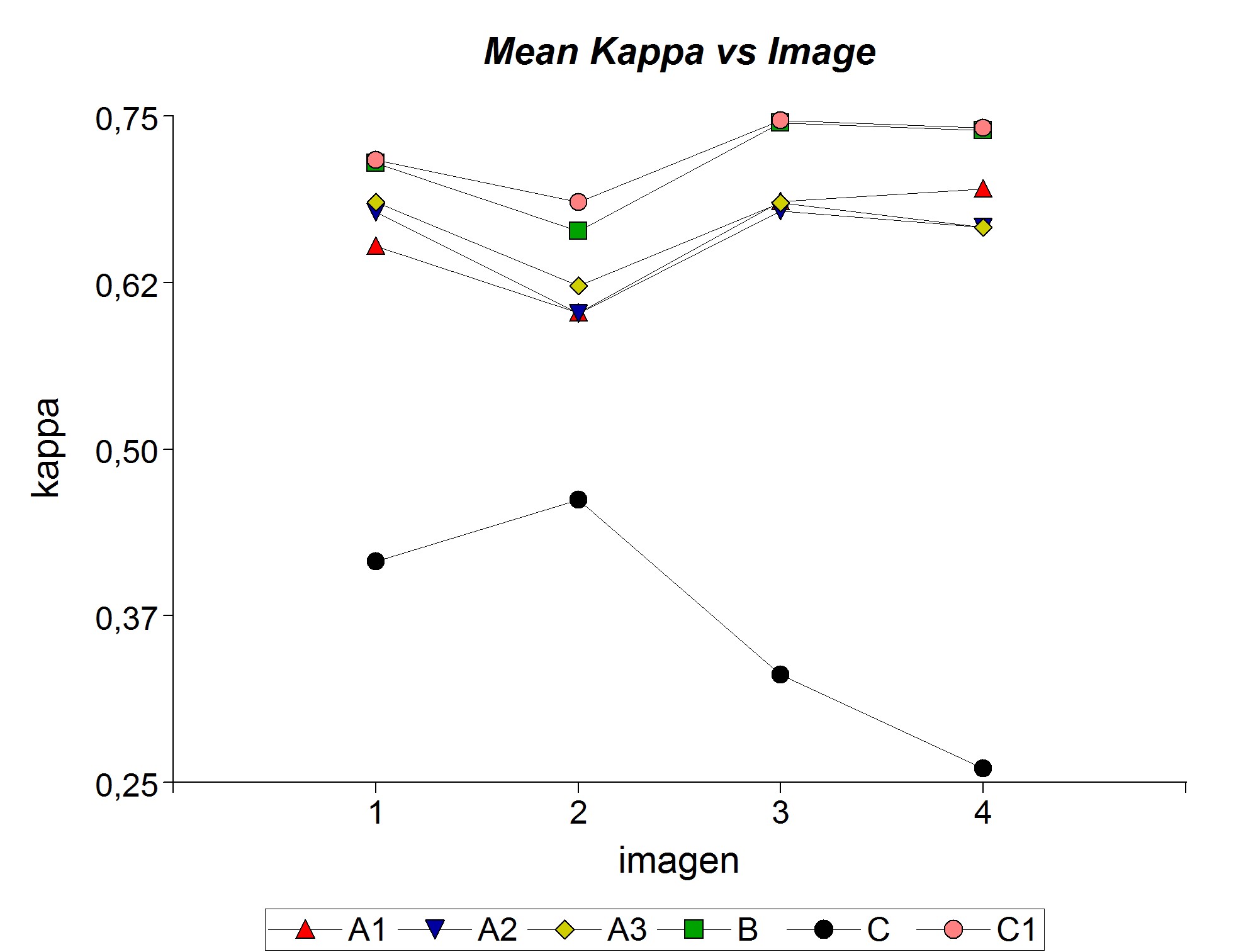}
\caption{Left panel :Overall Accuracy against Image,Right panel: Kappa against Image}
\label{classi} 
  \end{figure}

 In Figure \ref{classi} we see two profile plots, Mean Overall Accuracy against Image, and Mean Kappa against Image, where the means are made over all the sub-images of the same image.  The highest profile will be then the best method and the lowest the worse, if all profiles are more or less parallel.

In both plots, Method C performed badly, and seems to interact with the images. Its profile does not follow the pattern of the others. Nevertheless, Method C1, a simplification of method C, has good performance with these measures that consider good classification. It follows a similar pattern than Method A1,2,3 and B, but with scores as good as Method B.  Measures Kappa and Overall Accuracy depends heavily on good segmentation of the ground truth. Method C includes an enhancement that may give better defined classes, but separate it from the segmentation of the ground truth, increasing the global error, outside the imputation error.

\begin{figure}[h]
\centering
\includegraphics*[width=7.5cm]{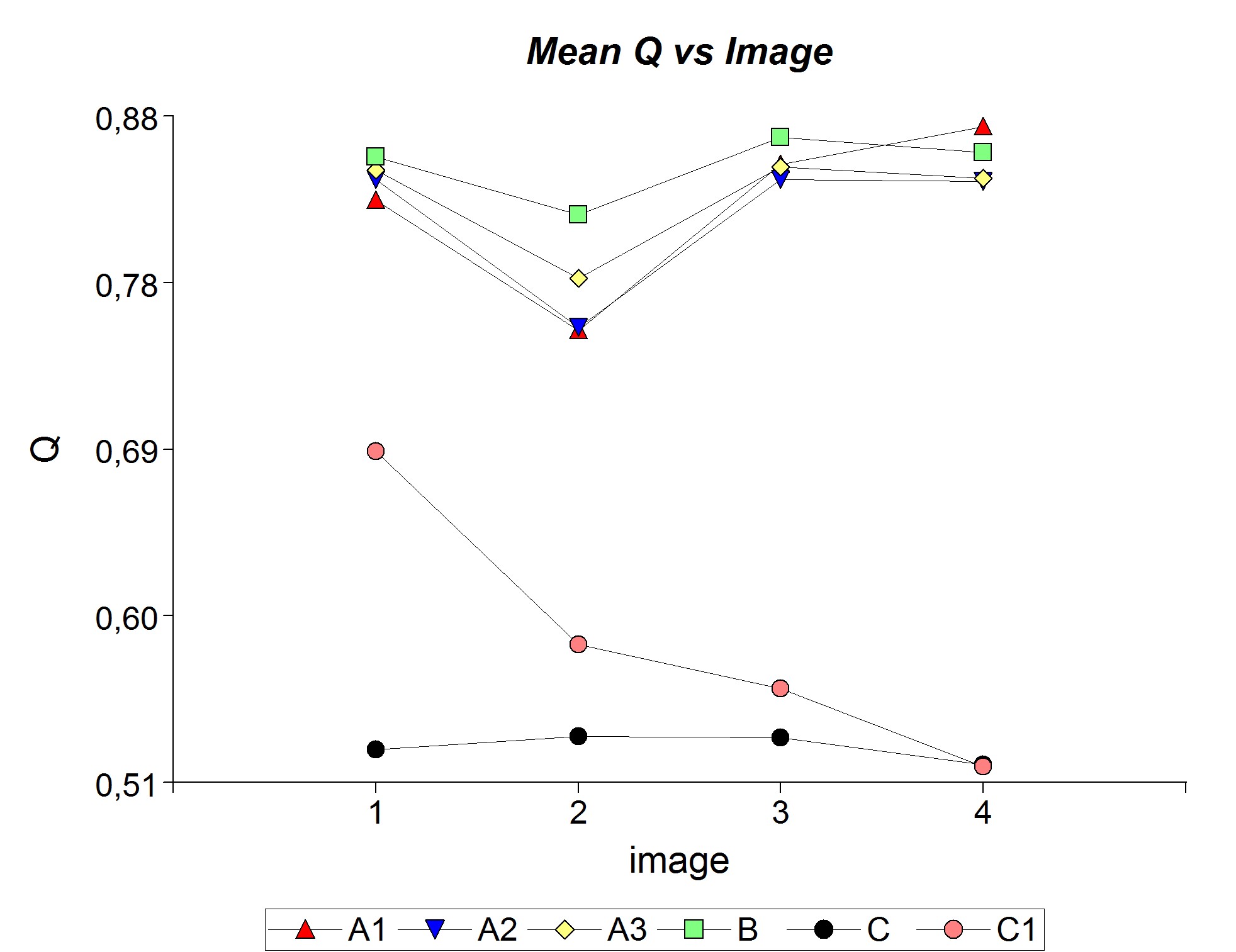}\includegraphics*[width=7.5cm]{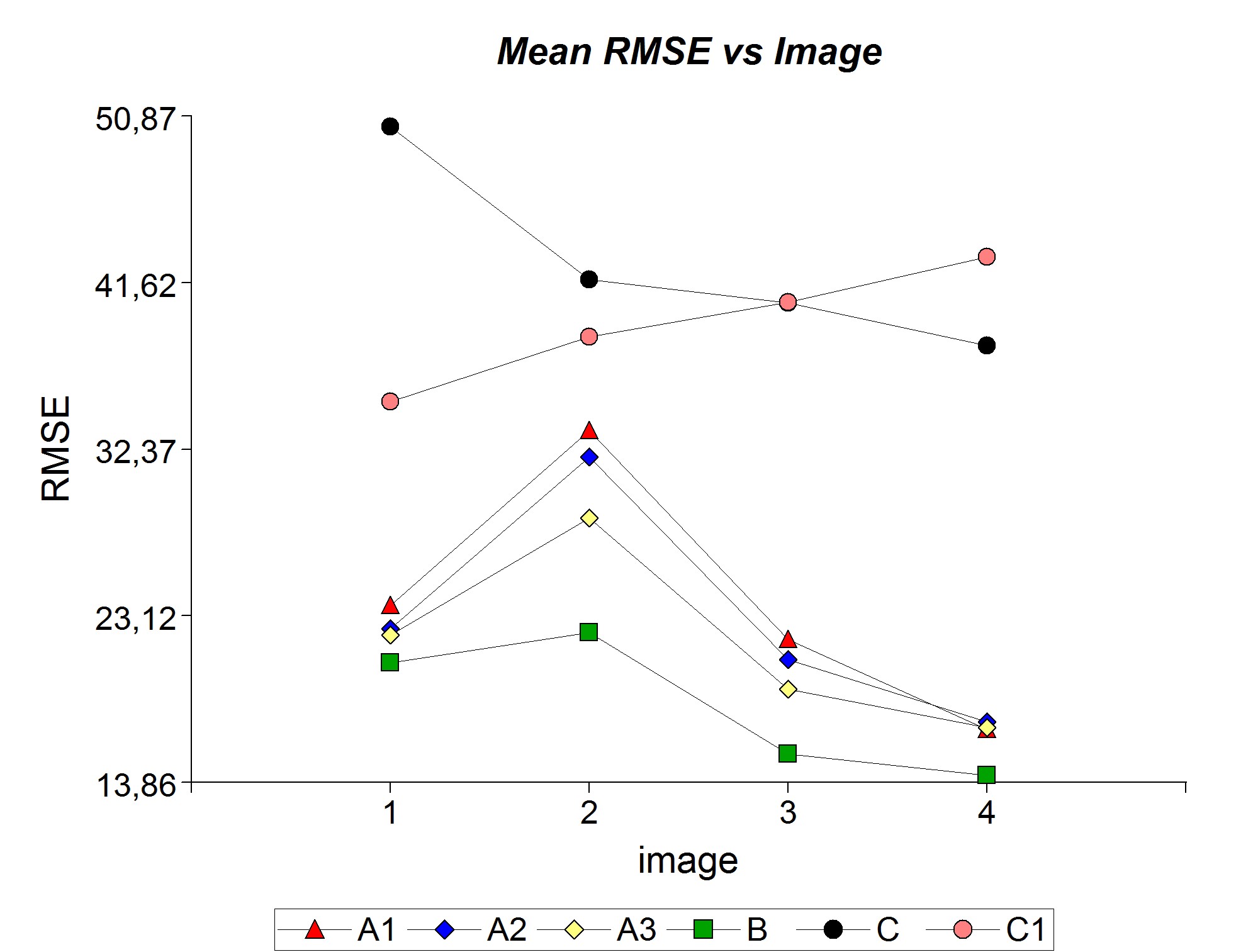}
 \caption{Left $Q$ agains Image and Right RMSE against image}
\label{radio}
      \end{figure}

 In figure \ref{radio} we see other two profile plots, Mean $Q$ against Image, and Mean RMSQ against Image. 

 The highest profile will be then the best method and the lowest the worse, when considering $Q$, and the reverse when considering RMSE,  if all profiles are more or less parallel. Again, Method C has bad performance, followed by Method C1. In this case, reasons for this behavior may be found in the fact that imputation is made randomly within each class, which could have a very wide range of radiometric values.  Method C1 may generate a map of classes very close to the real one, but not the radiometric values of the missing pixels. 

\begin{figure}[h]
\centering
\includegraphics*[width=7.5cm]{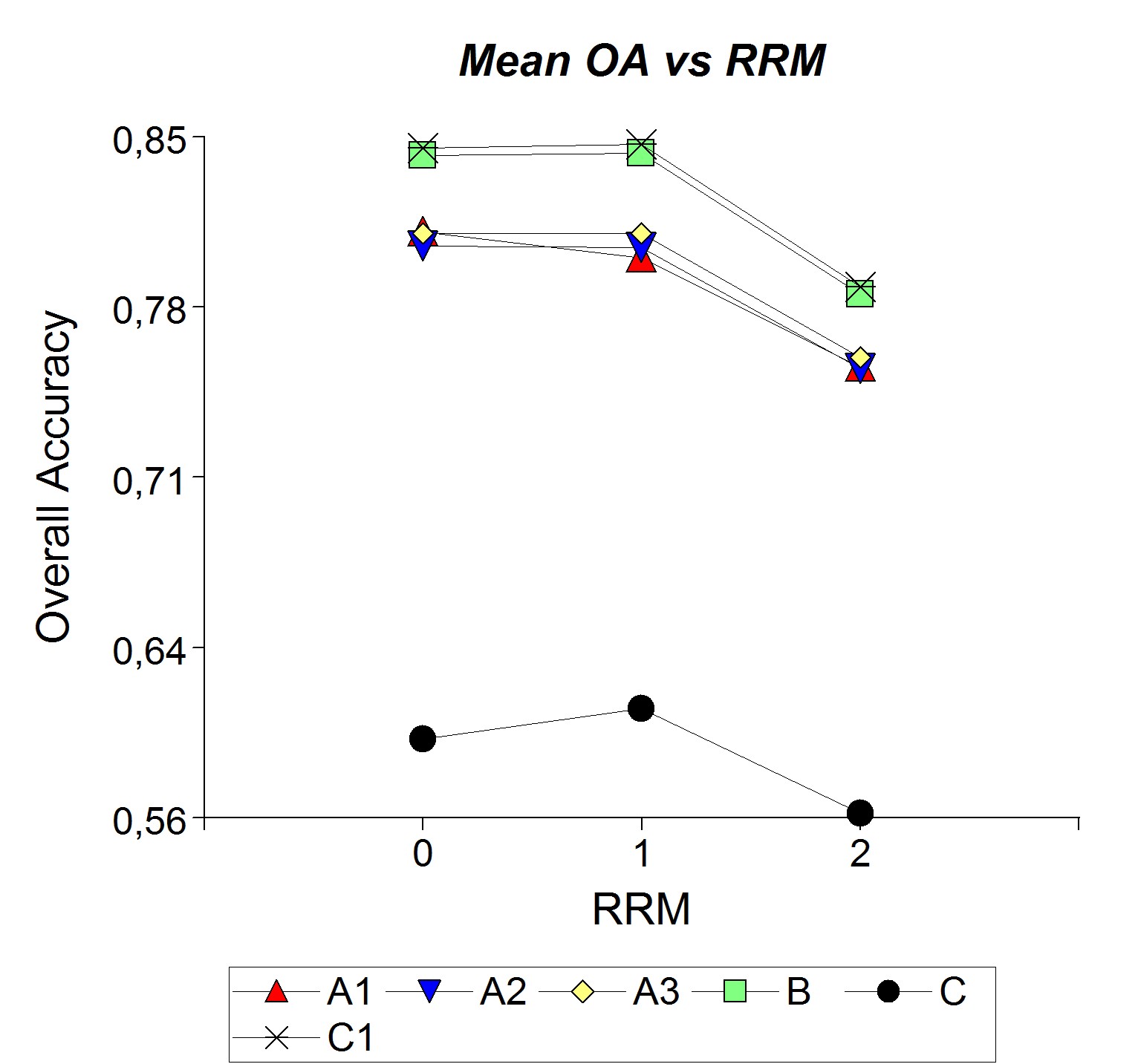} \includegraphics*[width=7.5cm]{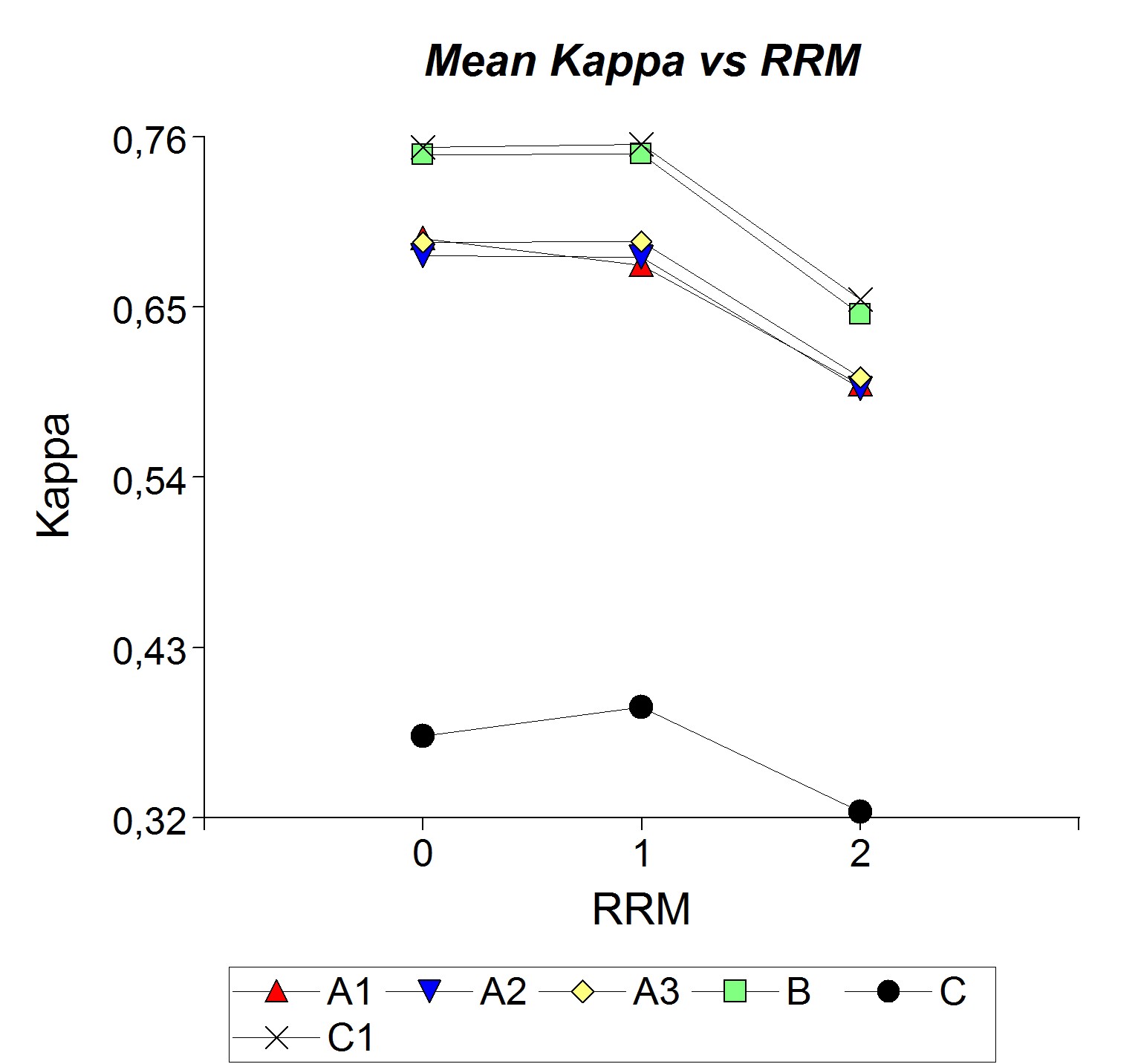}
\includegraphics*[width=7.5cm]{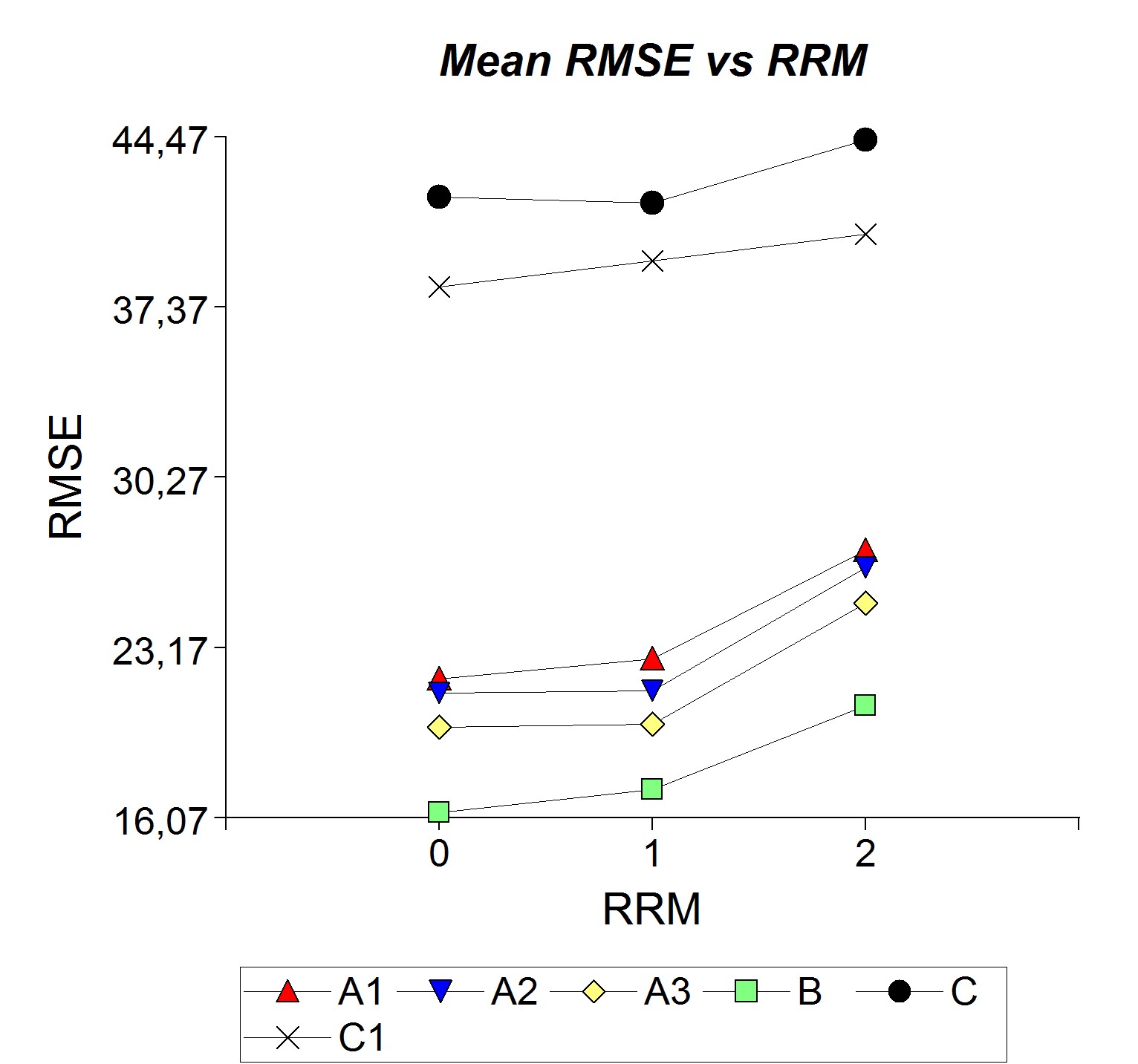} \includegraphics*[width=7.5cm]{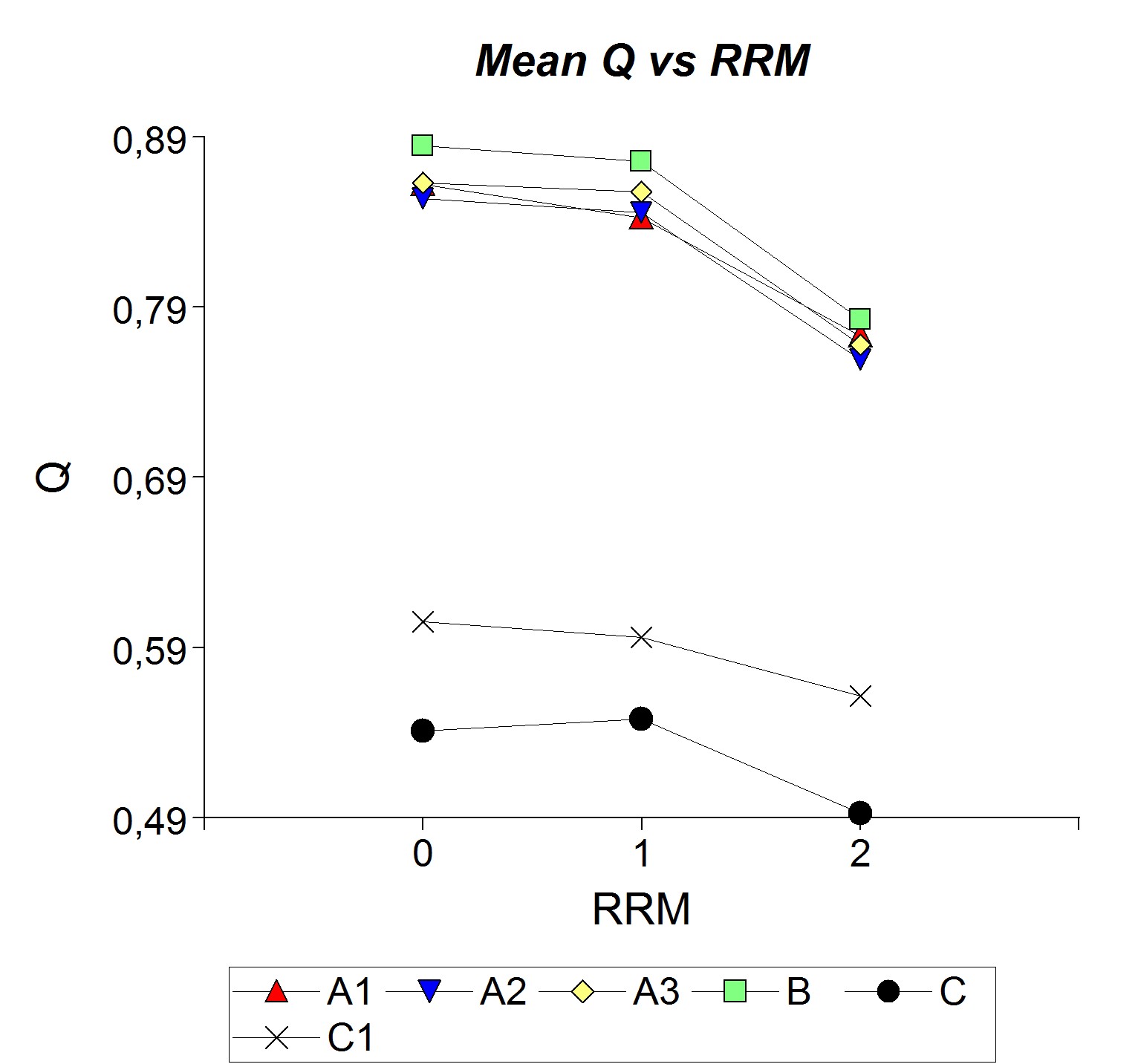}
\caption{From Left to Right. First Row: Overall Accuracy against RRM by  Method and Kappa against RRM by  Method. Second Row: RMSE against RRM by  Method and Q against RRM by  Method.   }
\label{RRM} 
  \end{figure}

We will now explore possible interactions between Imputation Method and Resolution Reduction Method (RRM), by making Mean Profile Plots of each Performance measure by Imputation Method and plotting it against RRM.  In Figure \ref{RRM} we should notice that the plots of three versions of Method A are not parallel, suggesting a possible interaction that should be further studied with an Anova test. 

\subsubsection{Preliminary conclusions}

In all cases, the mean profile of Method B seems to be best, followed by A3, fusion using a high value of threshold, that gives $Z$ more influence than $X_{old}$. That corroborates the facts pointed out by visual inspection on only one image. But we should test if the differences are statistically significant with level $\alpha=0.05$. All $A$ Methods are in pack, their differences may not be significant.

When studying interactions between the main factors, the $A$ Methods suggested a change between Mean performance using RRM 0 and RRM1. The most important change would be a significant difference on Mean Performance between Methods when using RRM3, which is slightly out of sync. 
 
In the next section, we will corroborate these empirical findings with Manova test of significance, and individual Anova tests. We will considering Image as a blocking factor, with a nested sub-factor, Sub-Image, discarding interaction as weak, Kuahl (1999). In both set of plots, we have reasons to believe that the profiles from Method C and C1 do not show interaction with the image factor, but other errors beside the ones measured by the particular performance index.

\subsection{Manova Test Outputs}

The vector of response variables contains the four measures considered, Overall Accuracy, Kappa, RMSE and Q. The Manova test search for significantly evidence to reject the following null hypothesis: 
\begin{equation*}
H1_{0}:\alpha _{i}=0,\ \forall i:i=1,..,6
\end{equation*}%
\begin{equation*}
H2_{0}:\gamma _{j}=0,\ \forall j=1,..,3.
\end{equation*}%
with $\alpha_i$ the vector of main effects introduced by the $i$- Imputation Method, and $\gamma_j$ is the vector of main effects introduced by the $j$- Resolution Reduction Method.  
Rejection of each hypothesis means

\begin{enumerate}

\item[$H1_0$]
There is a Imputation method effect, there is at least one method that has a different vector of performances than the others.

\item[$H2_0$]
There is a Resolution Reduction Method effect, there is at least one method of resolution reduction with a vector of performance different than the others.

\end{enumerate}
In a Manova test, answers are global, the combination of all performance measures give power to the rejection, Rencher (2002). Interaction between the main factors are important, since they could mask main effects. Following a Manova rejection, individual Anova test on each performance measure could give more information about relationships between the main factors, and individual comparisons may show possible clustering in Methods space.
\begin{table}[h]
\caption{Multivariate Analysis of Variance table (Wilks)}
\label{table1}
\begin{tabular}{|c|c|c|c|c|c|c|}\hline\noalign{\smallskip}
     			& S.V. &     		  F   Statistic&	df(num)&	df(den)	& p    	&      Error \\  \noalign{\smallskip}\hline\noalign{\smallskip}
Method    &     0,08	&200,56	& 	    20	&   3543	&$<$0,0001	 & \\               
RRM            & 0,64	  & 67,08	  & 	   8	  	 &2136	&$<$0,0001	        &   \\       
Image        &  	     0,61&	  2,57	&     12	   & 151	& 0,0040    & 	(im$>$subim)\\
image$>$subimage	&     0,01	& 36,78	&    240&	   4268	&$<$0,0001	                  &\\
Method*RRM      	&     0,95	&  1,52	&     40	  & 4052	 &0,0192	                  &\\ \noalign{\smallskip}\hline
\end{tabular}
\end{table}
Table \ref{table1} shows the results of the Manova test of significance with the Wilks statistics. Pillai, Roy an Lawley-Hotelling statistics were computed but leaded to the same conclusion and were omitted.  Interaction between Imputation Method and Resolution Reduction Method seems to be significant. Main effects are also significant. 

\begin{table}[h]\caption{Hotelling test (Bonferroni adjustment) Alpha:=0,05. Different letters indicate significant difference between location parameters (p$<$= 0,05)
Error: Pooled covariance matrix df: 1071}
\label{table2}
\begin{tabular}{|c|c|c|c|c|c|c|c|c|c|c|}\hline\noalign{\smallskip}
Method&	Overall Accuracy&	Kappa&	RMSE& 	 Q & 	n  	&  	 & 	& & &  \\  	\noalign{\smallskip}\hline\noalign{\smallskip}  
C     	    &  0,59	& 0,36&	42,67&	0,53	&192	&A 	 &	&& &   \\	  	  
C1    	    &  0,83 &	 0,72&	39,30&	0,59	&192	 & 	&B 	&  &&	  	  \\
B    	    &  0,83	& 0,71&18,09	&0,85  &192	  	  	&&&C 	& & 	\\  
A2    	   &   0,79	& 0,65&	23,05	&0,82	&192	  	  	  &&&&	D &	E \\
A1    	    &  0,79	& 0,66	&23,93&	0,83&192	  	&  	&&&  	D &	 \\ 
A3    	    &  0,80	& 0,66&	21,62&0,83	      &192	  	& &&&	  	  &	E \\ \hline\noalign{\smallskip}
\end{tabular}
\end{table}
In Table\ref{table2} we can see the output of a series of two means- Hotelling tests (with a Bonferroni adjustment for experiment-wise level $\alpha$=0.05), only with one factor, Imputation Method. Different letters indicate significant difference between location parameters. Using globally all measures, Method A2 seems no to be significantly different from Methods A1 and A3, but the last two are significantly different from each other. All other Methods are significantly different from each other.

\begin{table}[h]\caption{Hotelling test (Bonferroni adjustment) Alpha:=0,05. Different letters indicate significant difference between location parameters (p$<$= 0,05)
Error: Pooled covariance matrix df: 1069}
\label{table3}
\begin{tabular}{|c|c|c|c|c|c|c|c|}\hline\noalign{\smallskip}
RRM&	Overall Accuracy&	Kappa&	RMSE& 	 Q & 	n  	&  	  	& \\  	\noalign{\smallskip} \hline 		 \noalign{\smallskip} 
2  	   &  0,74	& 0,57	&30,71	&0,69	&384 	&A 	  & \\ 
1  	     & 0,79	& 0,66&	27,05	&0,76	&384	  	&  &B \\
0  	      &0,79&	 0,66&	26,58	&0,77&	384	  	&&  B \\\hline\noalign{\smallskip}
\end{tabular}
\end{table}

In Table \ref{table3}  we have written the output of a series of two means- Hotelling tests (with a Bonferroni adjustment for experiment-wise level $\alpha$=0.05), with the other Main Factor, Resolution Reduction Method. Different letters indicate significant difference between location parameters. Using globally all measures, there is a difference in performance  when there is lack of co-registration, while there is no difference between block averaging and CONGRID methods.

We have studied globally the main effects from Imputation and Resolution Reduction Methods. But interaction between both factors is not advisable. We would like to see in detail which one of the measures (if not all of them) shows changes in performance when both factors are considered at the same time.

\subsection{Individual Anova tests}

Knowing the results of the Manova test, we are confident that the level of the test of the individual Anova test are not far from its nominal value, $\alpha=0.05$.

In Table \ref{table5}, we have the output of a Anova test for Q as response variable.  Results are similar than Manova, and it shows indeed significant interaction between  Resolution Reduction Method and Imputation Method. The other three Anova Tests, one of each response measure, Kappa, RMSE and Overall Accuracy also indicate significant main effects, but non of them shows significant interaction between the two main factors, being $p$-values 0.54, 0.37 and 0.70 respectively for the interaction test. We display only Overall Accuracy output for the sake of completeness in Table \ref{table5}.
\begin{table}[h]
\caption{Analysis of variance table (Partial SS) Variable: Q measure and  Overall Accuracy}
\label{table5}
\begin{tabular}{|c|c|c|c|c|c|c|}\hline\noalign{\smallskip}
Q &      	 SS  	 &df 	& MS &	 F &   	p-value	&    (Error)\\    \noalign{\smallskip}\hline\noalign{\smallskip}
Model  &26,33&80&0,33	 &57,76&$<$0,0001	  &\\         
Method&19,85& 5&3,97	&696,52&$<$0,0001	&\\            
RRM    &1,48	&2&0,74&129,58&$<$0,0001	 &   \\            
Image & 0,51& 3&0,17	 & 2,36& 0,0803&	(Image$>$subimage)\\
Image$>$subimage&4,36&60&0,07&12,75&$<$0,0001	&\\                
Method*RRM&	 0,14 &10&0,01& 2,40& 0,0080	                &\\
Error   & 6,10&1071	&0,01& &&	\\      	       	                
Total   &32,44&1151& & &	&    	   \\  \hline\noalign{\smallskip}
\end{tabular}

\begin{tabular}{|c|c|c|c|c|c|c|}\hline
Overall Accuracy &      	 SS & 	 df 	  &MS &  	 F &   	p-value	&    (Error)     \\\hline
Model    &      	10,79	 &  80	 &   0,13 &	 43,07&	$<$0,0001	 &\\               
Method   &      	 7,72	 &   5	 &   1,54 &	493,21&	$<$0,0001 &\\          
RRM       &     	 0,70	 &   2	 &   0,35	 &112,28	&$<$0,0001 &\\            
Image     &     	 0,29	 &   3	 &   0,10	 &  2,77	& 0,0493&	(Image$>$subimage)\\
Image$>$subimage &2,06 &  60& 0,03 & 10,95&$<$0,0001&\\                
Method*RRM  & 0,02 &  10	&2,3E-03	 &  0,72	 & 0,7064	  &  \\              
Error        &  	 3,35 &1071	 &3,1E-03&&&	    \\  	       	                
Total&14,15&1151&&&&\\ \hline      	       	                
\end{tabular}
\end{table}
\subsection{Grouping}
We would like now to study how the location parameters cluster, to identify the methods that are significantly different from each other. We have some information form the multivariate Bonferroni corrected set of Hotelling test we made, but now we will see how the different variables cluster the Methods. 

We will use Fisher LSD multiple comparison test as clustering criteria. Other multiple comparison test like Tukey LSD and Duncan test gave the same type of clustering output.
 
\subsubsection{Grouping by Imputation Method}

We will study now the information contained on Table \ref{table6}.
\begin{table}[h]
\caption{Test: Fisher LSD, $\alpha=0,05$. Different letters indicate significant difference between location parameters.}
\label{table6}
\begin{tabular}{|c|c|c|ccccc|}\hline\noalign{\smallskip}
Method&	Mean OA	&n& & &&&\\ \noalign{\smallskip} 	  \hline	\noalign{\smallskip}  	  
C     	& 0,59&	192	&A & &&&\\	  	  
A2    	 &0,79&	192	&  &	B&&&\\ 	  
A1    	 &0,79&	192	&  &	B &&	& \\ 
A3    	 &0,80&	192	&  &	B 	&&& \\ 
B     	 &0,83&	192	&  &	  	&C&& \\
C1    	 &0,83&	192&  &	  	  	&C &&\\ \hline\noalign{\smallskip}
Method	&Mean Kappa&	n& & &&&\\  \noalign{\smallskip}	  	\hline \noalign{\smallskip} 	  
C    & 	 0,36&	192	&A 	  &&&&\\	  
A2   & 	 0,65&	192	 & 	&B 	  &&&\\
A1   & 	 0,66&	192	 & 	&B 	  &&&\\
A3   & 	 0,66&	192	 & 	&B 	  &&&\\
B    & 	 0,71&	192	 & 	  &&C&&\\ 
C1   & 	 0,72&	192	 & 	  	&&C&& \\ \hline\noalign{\smallskip}
Method	&Mean RMSE&n& & & &&\\  	  \noalign{\smallskip}	\hline  \noalign{\smallskip}	  	  	  	  	  
B     &  	18,09&	192	 &A & &&	 & 	  	  	  \\
A3   &  	21,62&	192	  &&B 	&&  	&  	  \\
A2   & 	23,05&	192	  &&B &	C &	  	  &\\
A1   & 	23,93&	192	  &&	&  C &	  	&  \\
C1   & 	39,30&	192	 & & & &	  	D 	  &\\
C    &42,67&192& & & &&E \\ \hline\noalign{\smallskip}
Method	&Mean Q&	n& & &&&\\  	  \noalign{\smallskip}	\hline 	  	\noalign{\smallskip}  	  	  
C    & 	 0,53	&192&	A &	  && & \\	  	  
C1   & 	 0,59	&192&	  &	B &	&  &\\	  
A2   & 	 0,82	&192&	  &	  &	C &&  \\
A1   & 	 0,83	&192&	  &	  &	C &&	  \\
A3   & 	 0,83	&192&	  &	  &	C &	 & \\
B    & 	 0,85	&192&	  &	  &	  &	D& \\ \hline \noalign{\smallskip}
\end{tabular}
\end{table}
When considering Imputation Method, Method B and C1 are grouped together by the two measures of good classification, Kappa and Overall Accuracy. The other two measures, Q and RMSE separate them into single clusters. The three versions of Method A have been grouped together by all the measures but RMSE, which considered Method A1 and A3 significantly different. The two versions of Method C, C and C1, have been distinguished by all the measures.
 
The multivariate simultaneous comparison distinguished all methods but the versions of Method A. And even though, Method A1 was considered significantly different from Method A3.  It is interesting to see that this is the way of grouping of the two radiometric measures. The classification based measures do not distinguish between Method B and C1, or A1, A2 and A3.

To the naked eye, Method B reconstruction of Figure \ref{method} is very different from Method C1 reconstruction, and Method A1 is sharper than Method A3. The multivariate simultaneous comparison agreed with this statement, with an experiment-wise Type I error $\alpha$=0.05. 

Now we can come back to the profile analysis of section \ref{profiles}. In that section, the performance measures means, computed over the sub-images of each image, were plotted as profiles, and the highest value of Q, Kappa and Overall Accuracy indicated the best Imputation Method, and the lowest value of RMSE back up the same statement. Method B seemed to be the best of all them. But the analysis did not have any statistical confidence. The simultaneous comparisons made reported Method B as significantly different in mean performance  from all the others, and Table \ref{table6} show the mean value of measures Q, Overall Accuracy and Kappa the highest of all, and the lowest of the error measure RMSE. These results give confidence to the previous profile analysis.

\subsubsection{Grouping by Resolution Reduction Method}

Now we are concerned with the fact that imputation may have reduced performance when the lower resolution image used as extra information is not co-registered accurately. When simulating the lower resolution image, it was chosen chose block averaging and CONGRID method as basic resolution reduction methods, and as a third method, the block averaged image was moved slightly (Method RRM2), simulating lack of co-registration. Simultaneous comparisons using Fisher LSD  (see Table \ref{table7}) report that the two versions of block averaging are indistinguishably, but RRM2  produced a significantly different mean in all measures, making them worse. In the case of Kappa, Q and Overall Accuracy, the means are smaller, and in the case of RMSE the mean is higher.

\begin{table}[h]
\caption{Test:Fisher LSD. $\alpha$:=0,05.  Different letters indicate significant difference between location parameters (p$<$= 0,05)}
\label{table7}
\begin{tabular}{|c|c|c|c|c|c|c|c|}\hline\noalign{\smallskip}	  
RRM&Mean OA&Mean Kappa& Mean RMSE&Mean Q&n& & \\	 \noalign{\smallskip}\hline\noalign{\smallskip}	  
2  &	 0,74&0,57&30,71   &0,69&		384	&A 	 &\\ 
0  &	 0,79& 0,66&27,05	&0,76&	384	&&  	B \\
1  &	 0,79& 0,66&26,58	&0,77&	384	&&  	B \\ \hline\noalign{\smallskip}	  
\end{tabular}

\end{table}

\section{Conclusions}

 Regression models are considered successful models for imputation in a wide range of situations in all Applied Sciences. In this paper, we introduced a Simple Regression Model for imputation of spacial data in large regions of a Remote Sensed image,Method B,  that  had a statistically significant better performance than two other main methods also adapted from the literature, in the frame of a careful simulation study.

Multivariate simultaneous comparisons of all imputation methods agreed with the visual inspection of the reconstructed images. Method B reconstruction shows less contrast between the imputed stripes and the non imputed regions, but looses sharpness, and appear slightly  blurred. Method A1 and Method A3 reconstructions are different, A1 produced more fine detail, but also a lot more contrast between imputed and non imputed regions, and A3 has a  smoother appearance, closer to Method B reconstruction, also with less detail.
 
 Method C and C1 were designed to give good segmentations, despite the large regions without informations, and Method C1 does. Pixels radiometric imputation is made choosing at random a a value form the convex hull of pixels of its class in a small neighborhood. Without the help of radiometric extra data, radiometric imputation become quite poor. Method C has a preprocessing step that makes the initial segmentation sharper, which increases the difference with the original image, not only on the imputed regions but in all regions.

 One of the hypothesis of all methods was the existence of good, co-registered, temporally accurate imagery with possible lower resolution. To test the dependence of the methods of co-registration, the block averaged image that act as lower resolution extra data was shifted slightly and the performance of all methods diminished. This reduction was observed statistically significant with Fisher LSD test of simultaneous comparisons, for each performance measure, and globally with a series of Hotelling tests (Bonferroni adjusted).

\section*{Acknowledgements}
The results introduced in this paper are part of the Masters Thesis in Applied Statistics of Valeria Rulloni, at Universidad Nacional de Cordoba.   This work was partially supported by  PID Secyt 69/08. We would like to thank S. Ojeda, J. Izarraulde, M. Lamfri, and M. Scavuzzo for interesting conversations leading to the design of the methods. The imagery used in the simulation section was  kindly provided by CONAE, Argentina.
Imputation methods and performance measures were computed with ENVI software. Statistical Analysis was made with INFOSTAT, UNC, provided by Prof. Julio Di Rienzo.

\end{document}